\documentclass[numbook]{svjour2}                    
\smartqed  
\usepackage{graphicx}
\usepackage{mathptmx}      
\usepackage[tbtags]{amsmath}
\usepackage{slashed}

\DeclareMathOperator{\sgn}{sgn}
\newcommand{\expectation}[1]{\langle#1\rangle}

\newcommand{\expectationfit}[1]{\left\langle#1\right\rangle}
\newcommand{\expectationtwo}[1]{\langle\langle#1\rangle\rangle}

\newcommand{\Or}{\mathrm{O}}

\newcommand{\rmi}{\mathrm{i}}

\begin{document}

\title{Wavefunction collapse via a nonlocal relativistic variational principle
}


\author{Alan K. Harrison}


\institute{Alan K. Harrison
\at Los Alamos National Laboratory, Los Alamos, NM   87544, USA
\email{alanh@lanl.gov}
}

\maketitle

\begin{abstract}
We propose, as an alternative theory of quantum mechanics, a relativistically covariant variational principle (VP) capable of describing both wavefunction collapse and, as an appropriate limiting case, evolution of the wavefunction according to the standard quantum mechanical (SQM) wave equation. This results in a nonlinear, nonlocal, time-symmetric hidden-variable theory; the hidden variable is the phase of the wavefunction, which affects the dynamics via zitterbewegung. 

The VP is
$\delta (A_1 + \epsilon A_2) = 0$,
in which $A_1$ and $A_2$ are positive definite integrals (over all spacetime) of functions of the wavefunction $\psi(t,\vec{x})$.
$A_1$ is quadratic in deviations of the wavefunction from compliance with the SQM wave equation.
$A_2$ is a measure of the uncertainty of the wavefunction, driving collapse by penalizing certain kinds of superpositions.
We also show that $A_1$ limits the rate of collapse, and that it enforces the Born rule, with suitable assumptions and approximations.

Since the VP optimizes a function $\psi$ of both space and time, the theory is not ``causal'' in the usual sense. Because it is not clear how Nature solves the optimization problem (e.g., whether a global or a local minimum is sought), we cannot yet say whether it is deterministic.
\end{abstract}

\section{Introduction}
Although the standard theory of quantum mechanics (SQM, for short) has been remarkably successful for many years, foundational and interpretational issues that troubled some of its founders (see e.g. \cite{EPR}) are by no means resolved. \cite{62_yrs,Open_Systems,Crossroads}
For instance, the wave equation is a time-symmetric, deterministic, linear differential equation. 
On the other hand,
the collapse process is not expressible as an evolution equation; the only prediction that this element of the theory can make
is of the probabilities of the various possible outcomes in an ensemble of identically prepared experiments.
The collapse mechanism is apparently unique, or almost so, among physical theories both because it is intrinsically random, and because it is asymmetric in time (separating a superposition of states in the past from a single state in the future).
In addition, the regimes of validity of the wave equation and the collapse process are defined in terms of whether or not a ``measurement'' is being performed, but that term (and the related term ``observer'') are not defined with the degree of clarity we expect
for such fundamental concepts. \cite{Bell_1990,Bell_Cosmologists}

The failure of SQM to give a detailed description of the collapse process is typically regarded as a conundrum for those who seek intuitive understanding, but not a practical weakness because it does not hinder calculations. However, one can design a set of thought experiments exhibiting the ambiguity of the theory. We can imagine an infinite set of experiments $\{E(\lambda)\}$ in which all the input parameters of the experiment are continuous functions of $\lambda$, with $0 \le \lambda \le 1$, such that $E(0)$ is clearly not a measurement and $E(1)$ is. For instance, if the measurement $E(1)$ requires turning on some probing electromagnetic field $F(t)$, then $E(\lambda)$ might be defined as the experiment conducted in the same way but using the field $\lambda F(t)$. SQM would say that $E(0)$ does not collapse the wavefunction and $E(1)$ does---so at what value of $\lambda$ does the collapse first take place, and how can one justify the abrupt transition from no collapse to collapse?

Because the results of the collapse process appear random---experiments with apparently identical initial conditions are observed to give
different results---it is natural to suspect that a ``hidden variable'' is involved, so that such a set of measurements in fact have different initial conditions. Then, presumably, when the hidden variable is accounted for, experiments that are indeed identically prepared
will yield identical results after all.
This promising resolution encounters an obstacle in Bell's Theorem 
\cite{Bell_1964} 
and relations equivalent to it \cite{Clauser_Horne_Shimony_Holt}, because 
that theorem implies that phenomena violating those inequalities cannot be
explained by local hidden-variable theories.

Because a great many alternative interpretations and theories of quantum mechanics have appeared in recent years, we will mention just a
few of the most successful.

Cramer \cite{Cramer_1986} has developed a ``transactional interpretation'' of QM that involves two-way ``transactions'' between lightlike separated points in spacetime. The spatial and temporal nonlocality of this approach enable it to explain nonlocal phenomena more plausibly than SQM can, in this author's opinion. However, since this is an interpretation and not a modification of the theory, the wave equation is unchanged, and the theory is
still unable to \emph{quantitatively} describe the transitions between measurement and non-measurement regimes (such as the dependence on $\lambda$ in the thought experiments described above). Cramer's interpretation may turn out to be consistent with the theory we will describe below, but we assert that it does not go far enough by itself.

Griffiths' ``consistent histories'' interpretation \cite{Griffiths_1984,Griffiths_2002,Thorndike} identifies time-ordered sets of mutually consistent events as ``histories.'' It enables the inference of probabilistic statements about unmeasured quantities, under appropriate circumstances. This is a very different approach than we have followed. Like Cramer's work, it is a re-interpretation but not a correction of the theory.

The collapse theories of Pearle, Ghirardi, Rimini and Weber \cite{Pearle_1979,GRW_1986,GPR_1990} propose that an external source
of noise (``hittings'') acts on the wavefunction in a way that produces the observed variety of experimental outcomes from a single initial condition. 
This results in some desirable properties, such as a much more rapid collapse for macroscopic objects than for very small systems. 
Our principal objection to this type of theory is that the external noise source lacks physical justification.

The decoherence explanation \cite{Zeh_1970,Zurek_1991,Tegmark_Wheeler,Decoherence_and_Classical_World} also relies on an external source of noise, in this case, the environment surrounding the system and the measurement apparatus. According to this view, unavoidable entanglement of these three regimes leads to the result that, when the system and apparatus are measured and the environment ignored, the system appears to have collapsed into a single state.
Although some questions have remained about this explanation,\cite{Schlosshauer} it is among the most successful attempts to explain the
collapse process.

As an alternative to relying on an external source of noise, and in search of a conceptually more compact theory, we will here consider a model which depends only on properties of the system under study, possibly including the measurement apparatus with which it is entangled.
This of course constitutes a hidden-variable theory; by the considerations mentioned above, it must be a nonlocal theory so as not to violate
the constraints of Bell's Theorem and the associated experimental findings. Although this program will lead to some unusual assumptions
about the operation of quantum mechanical systems in space and time, we will find that it holds great promise to explain issues and
observations that are troubling at best for other interpretations of quantum mechanics.

\section{Requirements to be satisfied by the theory}
\subsection{Constraints}
\label{Subsection:Constraints}
We intend the new theory to do the following things:
\begin{enumerate}
\item Describe matter waves themselves, and not principally the knowledge a human being actually or potentially has about the system under study. \item Describe wavefunction collapse (state reduction) in a natural way.\footnote{We do not require collapse to be \textit{instantaneous}. 
Under SQM, a decay time $\Delta t$ less than or of the order of $1/\Delta E$
is considered indistinguishable from zero, so the experimental evidence---which has been interpreted according to SQM---cannot rule out nonzero collapse durations satisfying that inequality. We therefore regard the term \emph{decay} as more accurate than \emph{collapse}, but will use the latter term in conformity with accepted nomenclature.}
\item Apply in the same way---that is, by solving the same equations---whether or not a measurement is being made (and thus not depend on the presence of an observer or the precise definition of ``measurement'' \cite{Bell_1990}).
\item Agree with the SQM wave equation under conditions in which the latter should hold (that is, when a ``measurement'' is not being made).
\item Predict experimental outcomes distributed according to the Born rule (for measurements typical of the body of experiments that have been done; we will explain that caveat presently).
\item Depend on a hidden variable or variables (allowable due to characteristic 11, below),
rather than an external ``noise'' source unrelated to the system under study,
to break the symmetry among possible experimental outcomes. 
\item Be deterministic (like other fundamental laws).
\item Be time-symmetric (like other fundamental laws).
\item Apply in the relativistic domain (and thus be expressible in covariant form). This requirement might seem excessive, because the issues we essay to address here are manifested in nonrelativistic SQM. However, we will propose below that the hidden variable central to the choice of outcomes of the collapse process is the phase of zitterbewegung oscillations, for which the simplest description is relativistic. Therefore we find it necessary to construct a relativistic theory to explain experimental observations that, in every other way, are completely confined to the nonrelativistic domain. We are of course hopeful that the relativistic theory outlined below will apply as well to explain fully relativistic phenomena (although in this initial work we have stopped short of a field-theoretic analysis, which may well be needed at some point). But for the time being, to keep the scope of this paper somewhat bounded, we will focus our attention on measurements and \emph{gedanken} experiments at low energies and very subluminal speeds.
\item Be nonlinear. This follows from 
conditions 2 and 3, which mean that the collapse must follow from the fundamental formulation (e.g., wave equation) of the theory, without the need for auxiliary constructions or variables (like de Broglie-Bohm pilot waves \cite{Bohm_1952a,Bohm_1952b}).
\item Be nonlocal in space 
both to be able to distinguish pure eigenstates from superpositions,
and to avoid the Bell's Theorem \cite{Bell_1964} prohibition on local hidden-variable theories (cf. condition 6).
\cite{Mermin_1993}
\item Be nonlocal in time. This is implied by conditions 11 and 9, because observers in different reference frames must agree on whether the wavefunction at a spacetime point $A$ depends on conditions at $B$, but they may not agree on whether $A$ and $B$ have the same time coordinate.
\item Cross over smoothly and naturally from microscopic to macroscopic behavior, dispensing with the need for two sets of rules in regimes separated by a boundary (the ``shifty split'' \cite{Bell_1990}). 
\end{enumerate}

Obviously, we must explain how a new theory that differs from SQM can be consistent with the experimental record, which is generally understood as being consistent with SQM. Therefore, in pursuit of that explanation, discussions below of experimental and measurement processes are meant to refer to processes and technologies that are likely to have been employed up to the present time, unless otherwise indicated. For instance, the experimental record is generally understood to be consistent with the Born rule, and the theory to be presented below will explain that in terms of standard experimental techniques of the past and present, but will not rule out the possibility that future experiments may contradict the Born rule by detecting or manipulating the hidden variable, or by using technology not generally available up to now. [This is the reason for the caveat appended to constraint 5 above.]
\subsection{Phase as the hidden variable}
We propose that the hidden variable is the phase of the wavefunction at some time relative to the measurement
(this choice was explored by Pearle \cite{Pearle_1976})---
or equivalently, the start time of the experiment relative to the oscillations of the wavefunction. Suppose we write an energy eigenstate $j$ as
\begin{equation*}
\psi_j = e^{-i E_j t} \chi_j(\vec{x}) \, .
\end{equation*}
(We will take $c \equiv \hbar \equiv 1$.)

If $\psi$ is a superposition of states with different energies
\begin{equation*}
\psi = \sum C_j e^{-i E_j t} \chi_j
\, ,
\end{equation*}
then there exist operators $\mathcal{O}$ such that the combination $\psi^\dagger \mathcal{O} \psi$ contains cross terms (``beats'') that oscillate in time 
\cite{Thaller}. 
For instance, if $\mathcal{O}$ commutes with functions of time,
\begin{equation}
\label{beats}
\psi^\dagger \mathcal{O} \psi
= \sum_{j,k} \left[ C_j^* \, C_k \, e^{i(E_j - E_k)t} \, \chi_j^\dagger \, \mathcal{O} \, \chi_k 
+ C_k^* \, C_j \, e^{i(E_k - E_j)t} \, \chi_k^\dagger \, \mathcal{O} \, \chi_j \right] \, .
\end{equation}

Experimental detection and measurement of the wavefunction require it to be localized, at some stage of the experiment, to within a spatial extent comparable to the size of the laboratory. This means that the localized wavepacket includes contributions from negative-energy modes \cite{BD,Strange,Greiner},
so the sum \eqref{beats} includes terms (zitterbewegung) for which $E_j$ and $E_k$ have opposite signs.

Now suppose that an experiment is begun at some time $t_{\rmi}$, and that an experimental result is read at some time $\tau$ after the beginning of the experiment. \footnote{Actually, the ``reading'' of the result is likely to be a process that takes place over a range of times, not a single instant, but that additional complication does not affect the conclusion that we will reach.} Substituting 
\begin{equation}
\label{define_tau}
t=t_{\rmi}+\tau 
\end{equation}
into \eqref{beats} gives an expression that varies sinusoidally with $t_{\rmi}$, with a period $T_{jk} = 2 \pi / (E_j - E_k)$ in each term of the sum. The experimenter typically controls $\tau$ but would not attempt to control $t_{\rmi}$, both because he/she is ignorant of the initial phase of the wavefunction and because, for the zitterbewegung terms, $T_{jk}$ will not significantly exceed
\begin{equation}
\label{very_short_period}
\pi / m \simeq 4 \times 10^{-21} \text{ seconds for electrons} \, ,
\end{equation}
so control of $t_{\rmi}$ to within a fraction of $T_{jk}$ is not currently attainable. 
Therefore a set of ``identically prepared'' experiments would be expected to have different values of $t_{\rmi}$ and hence different outcomes.
For this reason we can regard $t_{\rmi}$ as the hidden variable. To very good approximation, we regard it as a random variable chosen from a uniform distribution on $[0,T]$ for some $T \gg T_{jk} \thickspace \forall j,k$.

Since the zitterbewegung terms in $\psi^\dagger \mathcal{O} \psi$ are likely to vary from one experimental realization to another, for the reasons just given, we propose that they determine the outcome of the measurement. Since nonrelativistic systems do not contain appreciable contributions from negative-energy states, the zitterbewegung terms are small in such systems. It is natural to ask whether they are large enough to drive wavefunction decay, or at least to determine the outcome of such decay.

Consider a nonrelativistic wavepacket localized in space, which therefore contains some negative-energy components. If it is confined to a size $\Delta x$, then its representation as a superposition of plane waves must include contributions from modes with momenta up to $\Delta p \simeq 1/\Delta x$, and therefore the negative-energy states have an amplitude $C_j$ of order $\Delta p/m \simeq 1 / m\Delta x$ relative to the dominant positive-energy states. \cite{BD}

Now we would expect one lower bound on the time required for a measurement to be $\Delta t \simeq \Delta x$. If the zitterbewegung terms are to drive the wavefunction decay, their strength must therefore scale inversely with the size of the system, in order for the decay to be essentially complete on an experimental timescale. But that is what we have just shown.

Thus we expect that \textit{even in nonrelativistic systems}, zitterbewegung terms drive the decay to a single state when a measurement is made. Therefore the theory we seek must be a relativistic one.
\section{Formulation of the variational principle}
Because some of the constraints we hope to satisfy (decay to a single state, Born rule) are stated in terms of the eigenstates of the measured variable, as defined in SQM, and because we expect SQM to emerge as a special case of our theory, we will develop it in terms of the equations of SQM and their solutions.

We will focus on nonrelativistic systems to develop the theory, even though we have found that a relativistic theory will be needed for that task. (We would like this theory to apply to relativistic systems as well, but there are sufficiently many interesting problems at low energies that we can focus our attention on them for the time being.) Thus we will consider only cases for which the energy changes and differences due to the experimental process (the imposed fields), and the rates of change of those energies and those fields, are small compared to the rest energy $mc^2$ of the particle.
In this paper we limit consideration to fermions as the system to be described, so the relevant SQM wave equation is the Dirac equation 
\cite{BD,Strange,Greiner} 
\begin{equation}
\label{Dirac}
\mathcal{D} \psi = 0 \, ,
\end{equation}
where $\mathcal{D} \equiv \slashed\pi/m - 1$.

If we use the representation
\begin{equation}
\label{choice_of_gammas}
\gamma^{\,0} = {
\begin{pmatrix}
1  & \,\,0 \\
0  & -1
\end{pmatrix}
}, \qquad
\gamma^{\,i} = {
\begin{pmatrix}
0  & \sigma^i \\
-\sigma^i  & 0
\end{pmatrix}
}
\end{equation}
of the $\gamma$ matrices, then positive-energy wavefunctions  of momentum $p \ll m$ look like
\begin{equation}
\label{order_magnitude_positive}
\psi \propto {
\begin{pmatrix}
\Or \, (1) \, \\
\Or \, (1) \, \\
\Or \! \left(\frac{p}{m}\right) \\
\Or \! \left(\frac{p}{m}\right)
\end{pmatrix}
}
\end{equation}
and negative-energy wavefunctions like
\begin{equation}
\label{order_magnitude_negative}
\psi \propto {
\begin{pmatrix}
\Or \! \left(\frac{p}{m}\right) \\
\Or \! \left(\frac{p}{m}\right) \\
\Or \, (1) \, \\
\Or \, (1) \,
\end{pmatrix}
}
\end{equation}

\subsection{Terms in the variational principle}
\label{Subsection:VP_terms}
As explained above, the desired theory must be nonlocal in space and time; therefore a natural mathematical form would includes (an) integral(s) over spacetime. Thus the wave equation must be an integral or integrodifferential equation. This suggests in turn that we formulate the theory as a variational principle (VP).

We propose as such a principle that nature seeks to minimize the sum of two positive definite terms: one that vanishes when the wavefunction is a solution of the SQM wave equation, and so tends to drive the wavefunction toward such solutions; and one that increases with the uncertainty in the wavefunction, thus favoring minimum uncertainty states 
\cite{Schwinger}.

A superposition of eigenstates of the operator corresponding to the measured property will generally have more than the minimum uncertainty. Therefore the tendency toward minimum uncertainty states will tend to cause such superpositions to decay, as we expect when measurements are made. This idea will be illustrated below, in the example calculation of the two-slit experiment.

The desired VP takes the form
\begin{equation}
\label{variational_principle}
\delta (A_1 + \epsilon A_2) = 0
\end{equation}
in terms of positive definite functionals $A_1$ and $A_2$ of the wavefunction and a positive dimensionless constant $\epsilon$. $A_1$ measures the deviation of the wavefunction from a solution of the SQM wave equation, so minimizing it drives the wavefunction to obey that equation.
We will see that $A_1$ plays two other roles: it forces the collapse to take place over a period of time rather than instantaneously, and it enforces the Born rule.

The second term, $A_2$, is a measure of the position-momentum (and energy-time) uncertainty of the wavepacket.
We propose that under conditions typical of a measurement,
minimizing this term drives reduction of the wavefunction to a single eigenstate of the operator corresponding to the quantity being measured.
Since $A_2$ must somehow select a state pertaining to that particular operator, we must include the measurement apparatus
(or some part of it that is entangled with the system being measured)
in the wavefunction that appears in the VP; then the tendency of $A_2$ to minimize wavefunction uncertainty will make it unlikely for a
measurement involving a macroscopic apparatus to end up in a superposition of macroscopic (``pointer'') states.

For an example in which part of the measurement apparatus is included in a VP calculation, see this author's calculation \cite{Harrison_FQMT}
of the electron two-slit experiment.

The positive dimensionless number $\epsilon$ in the variational principle \eqref{variational_principle} allows us to adjust the relative sizes of its terms. We shall take it to be a constant, although in a more elaborate theory it could depend on $\psi$ in some way. 
Its magnitude is unknown at present, but we may be able to measure or infer it in the future, as we gain more experience with the VP.

$A_1$ and $A_2$ are in general integrals over all of spacetime, but the variational principle can still be useful without solving for the entire history of the universe. This is because an experiment can usually be considered to be localized within some region $\mathcal{R}$ of spacetime, meaning that states and events within $\mathcal{R}$ do not interact with the exterior (complement) of $\mathcal{R}$.
\footnote{This is of course an idealization, since the experimenter must interact with the experiment to set it up and to read out the measurement.
The implications of this for the application of the VP and the definition of $\mathcal{R}$ are a matter for further study.}
Since we expect the system to evolve according to the SQM wave equation except when a measurement is being made, the spacetime regions surrounding those measurements are islands surrounded by regions in which the SQM wave equation is satisfied and the integrands  in $A_1$ and $A_2$ take their minimum values. We conjecture that the minimization problem for all spacetime then reduces to the problem of minimizing $A_1 + \epsilon A_2$ over each such ``island,'' and that we can do so for each island independently of all the others. As a result, the global variational principle \eqref{variational_principle} can be reinterpreted to apply to a single island. We conjecture that in a properly designed experiment, the experimental domain $\mathcal{R}$ includes all of one (or more) island(s), that is, that its boundaries include only spacetime points where the SQM wave equation is satisfied (to a degree of accuracy commensurate with the requirements of the experiment).

Therefore we will write $A_1$ and $A_2$ as integrals over all of spacetime,
with the understanding that it is generally permissible to limit the domains of integration to some bounded region $\mathcal{R}$.

\subsection{$A_1$ term---preference for solutions of the SQM wave equation}
We will take as the first term in the VP
\begin{equation}
\label{define_A_1}
A_1
\equiv
\expectationfit{ \expectationfit{ \mathcal{D}^\dagger \mathcal{D} }}_1
.
\end{equation}
Here our notation $\expectation{\expectation{\mathcal{O}}}_1$ signifies the expectation value of an operator $\mathcal{O}$
that depends on a single spacetime coordinate $x^{\mu}$ per particle.
If we are dealing with single-particle states $\psi$, this is defined in the usual way:
\begin{equation}
\label{define_<O>}
\expectationfit{\expectationfit{\mathcal{O}}}_1 \,
\equiv \,
\frac
{\int d^4 x \,\, \psi^\dagger(x) \, \mathcal{O}(x) \, \psi(x)}
{\int d^4 x \,\, \psi^\dagger(x) \, \psi(x)}
\, .
\end{equation}
The double triangular brackets are to distinguish this notation from the three-dimensional matrix element
\begin{equation*}
\left\langle \psi | \mathcal{O} | \eta \right\rangle_t 
\equiv
\int d^3 x \,\, \psi^\dagger(t, \vec x) \, \mathcal{O}(t, \vec x) \, \eta(t, \vec x) \, ,
\end{equation*}
which will be useful to us later on.
(A useful mnemonic is that a single pair of brackets stands for an average over space alone, while a two pairs signify averaging over both space and time.)
Then the one-point expectation could be written as
\begin{equation*}
\expectationfit{\expectationfit{\mathcal{O}}}_1 \,
=
\frac
{\int dt \left\langle \psi | \mathcal{O} | \psi \right\rangle_t }
{\int dt \left\langle \psi | 1 | \psi \right\rangle_t }
\, .
\end{equation*}
We see that as expected, this form of $A_1$ penalizes deviations from the Dirac equation; 
solutions of that equation trivially give $A_1$ its minimum value of zero. In fact, there is no other way to get
$A_1 = 0$, so the SQM wave equation $\mathcal{D} \psi = 0$ is both a sufficient and a necessary condition for $A_1$
to vanish.

We believe that it should be possible to construct analogous VPs for other SQM equations, such as the Klein-Gordon equation,
by making appropriate substitutions for the operator $\mathcal{D}$ in \eqref{define_A_1},
but as mentioned before, this paper is limited to fermions.
\subsection{Small-perturbation case}
It will be useful to write the wavefunction as a superposition of modes, and because we will
need to understand how the superposition evolves in time, we will now consider ``small-perturbation''
limitations under which such modes can be defined and followed in time. These limitations will help us
understand the VP, but we do not mean to imply that its validity is limited to this case
(or to a fixed reference frame, in which we will do the following analysis).

Later in the paper, we will consider fidelity of our proposed quantum mechanical principle to the Born rule.
Since experimental tests of the Born rule must relate the measurement outcomes
to the initial structure of the system under study, the measurement must be made in a way that
perturbs the system only slightly.
A large perturbation would ruin that relationship---but (generally speaking) a perturbation-free measurement is impossible.

We will find it useful to write the Dirac equation in the form
\begin{equation}
\label{Dirac_like_Sch}
i \frac{\partial\psi}{\partial t} = H\psi
\end{equation}
where the Dirac Hamiltonian $H$ is defined so as to make \eqref{Dirac_like_Sch} equivalent to \eqref{Dirac}:
\begin{equation*}
H = \gamma^{\,0} (\vec \gamma \cdot \vec \pi + m) + e A^0.
\end{equation*}
Now the state energies will be important to our analysis of zitterbewegung, so we would like to express them as the eigenvalues of the Hamiltonian. Thus we will supplement the Dirac equation \eqref{Dirac_like_Sch} with a Schr\"odinger-like eigenvalue equation
\begin{equation}
\label{energy_eigenvalue_eqn_1st}
H \psi = E \psi
\end{equation}
and look for eigenstates that solve both equations.

Since any experiment is carried out by manipulating the Hamiltonian, 
the latter must be a function of $\tau$, via the electric and magnetic potentials $\Phi$ and $\vec A$, but not otherwise on $t$, because it contains only spatial derivatives.
(Recall that the experimenter is ignorant of $t_{\rmi}$.)
Therefore $\tau$ can be held fixed and the equation solved as a function of the spatial coordinates. As a result, for any value of $\tau$, the Hilbert space is spanned by a basis of eigenfunctions $\chi_j(\tau, \vec x)$:
\begin{equation*}
H(\tau, \vec x) \, \chi_j(\tau, \vec x) = E_j(\tau) \, \chi_j(\tau, \vec x)
\end{equation*}
satisfying the orthonormality relation in 3-space
\begin{equation}
\label{orthonormality_xi}
\int d^3 x \,\, \chi_j^\dagger(\tau, \vec x) \, \chi_k(\tau, \vec x) = \delta_{jk} \, .
\end{equation}

Let us suppose that $H$ varies continuously with $\tau$. Therefore we expect that for any $j$, $E_j(\tau)$ and $\chi_j(\tau, \vec x)$ are also continuous functions of time $\tau$, except possibly for a set of measure zero of values of $\tau$; let us assume that the experiment is designed so that those special cases are not encountered. Put another way, we number the eigenstates $\chi_j(\tau+\delta \tau, \vec x)$ in a way consistent with our numbering of them at $\tau$, so that 
\begin{equation*}
\lim_{\delta \tau \to 0} \chi_j(\tau + \delta \tau, \vec x) = \chi_j(\tau, \vec x) \, .
\end{equation*}
So a given state (choice of $j$) retains its identity as time evolves.

Now we are in a position to be more precise in our statements that the perturbations in the Hamiltonian are small. Since each state is normalized to unity [Eq. \eqref{orthonormality_xi}], we require the rates of change to be small, that is,
\begin{equation}
\label{slow_varying_xi}
\int d^3 x \, \chi_k^\dagger(\tau, \vec x) \, \frac{\partial}{\partial \tau} \, \chi_j(\tau, \vec x) \ll m
\end{equation}
for any choices of $j$, $k$ and $\tau$.

We have constructed the basis set $\{\chi_j\}$ as solutions of the eigenvalue equation \eqref{energy_eigenvalue_eqn_1st}, but not the Dirac equation \eqref{Dirac_like_Sch}. To get a basis that satisfies that equation as well, we note that any solution $\chi_j$ of the eigenvalue equation will still be a solution if it is multiplied by a function of $t$. Therefore
\begin{equation}
\label{define_psi_basis}
\psi_j(t, \vec x ; t_{\rmi}) \equiv \chi_j(t-t_{\rmi}, \vec x)
\exp \left[ -i \int_0^t E_j(t^\prime - t_{\rmi}) \, dt^\prime \right]
\end{equation}
is a solution of the eigenvalue equation. (Note that we lose no generality by choosing $t^\prime =0$ as the lower limit of integration, since the choice of the origin of time has not been, and will not be, otherwise constrained; note also that that equation is acceptable for all real values of $t$, not just $t \ge 0$.) Our notation signifies that $\psi$ is the function of $t$ and $\vec x$ that results from starting the experiment at time $t_{\rmi}$. The dependence on $t_{\rmi}$ will become important later on, but for the time being we will omit it from our notation for $\psi$.
Then
\begin{equation*}
i \frac{\partial}{\partial t} [\psi_j (t, \vec x)]
= E_j(t-t_{\rmi}) \psi_j (t, \vec x)
+ i \left. \frac{\partial \chi_j}{\partial \tau} \right| _{\tau = t-t_{\rmi}}
\mspace{-18mu} \exp{\left[ -i \int_0^t E_j(t^\prime - t_{\rmi})dt^\prime \right]} .
\end{equation*}
Now by the assumptions laid out above, the second term on the right-hand side is negligible compared with the first term; we can see this by operating on those two terms from the left with the operator
\begin{equation*}
\int d^3 x \, \chi_k^\dagger(\tau, \vec x)
\end{equation*}
for all possible basis states $\chi_k$, using \eqref{slow_varying_xi}, and remembering that for nonrelativistic potentials, the Dirac equation gives $|E| \approx m$. Then since $\psi$ satisfies the eigenvalue equation, we can write
\begin{equation*}
i \frac{\partial}{\partial t} [\psi_j (t, \vec x)]
= H(t-t_{\rmi}, \vec x) \psi_j (t, \vec x),
\end{equation*}
so the states $\{\psi_j\}$ satisfy the Dirac equation as well as the eigenvalue equation.
This property will allow us to easily switch between the eigenvalue equation, which we will use in our analysis of the Born rule, and the relativistic expressions and equations that appear in the covariant form of the theory. 

Note also that equation
\eqref{define_psi_basis} almost completely separates the space and time dependence of $\psi_j$, 
since $\chi_j$ depends on position but only slightly on time (via $\tau$). Because of our earlier stipulations on small and slow perturbations of the Hamiltonian, $\chi_j$ is to leading order a function of position alone. However, the fact that the relative change of $E_j$ is also very small does not mean that it can be ignored in the exponential, as we have no reason to presume that
\begin{equation*}
\int_{t_0}^t \left[E_j(t^\prime- t_{\rmi}) - E_j(0) \right] \, dt^\prime \ll 1 \, .
\end{equation*}

Now the set $\{\psi_j(t)\}$ are a complete basis satisfying
\begin{equation}
\label{psi_are_Dirac_solutions}
\mathcal{D} \psi_j = 0
\end{equation}
and
\begin{equation}
\label{orthonormality_psi}
\int d^3 x \, \psi_j^\dagger(t, \vec x) \, \psi_k(t, \vec x) = \delta_{jk}
\, ;
\end{equation}
we can therefore expand a general wavefunction $\psi$ as
\begin{equation}
\label{expansion_in_normal_modes}
\psi(t, \vec x) = \sum_j C_j(t) \, \psi_j(t, \vec x)
\, .
\end{equation}
\subsection{$A_1$ term---penalty for rapid evolution of the wavefunction}
The second property of $A_1$ is easily understood if we expand the wavefunction
in terms of eigenfunctions of the SQM (Dirac) operator as in \eqref{expansion_in_normal_modes}.
Then from \eqref{psi_are_Dirac_solutions},
\begin{equation}
\label{identity_pi-m}
\mathcal{D} \psi
= \frac{i \gamma^{\,0}}{m} \sum_j C_j^{\, \prime} \, \psi_j
\, ,
\end{equation}
and it follows immediately that
\begin{equation}
\label{A_1_as_sum_of_C_prime_sqd}
A_1(t) = \frac{
\int \! dt \, \sum_j \lvert C_j^{\, \prime}(t) \rvert^2
}
{
m^2 \int \! dt
}
.
\end{equation}
Thus $A_1$ penalizes rapid changes in the coefficients $\{ C_j \}$; 
for instance, instantaneous collapse ($\lvert C_j^{\, \prime}(t) \rvert \propto \delta(t)$) would make an infinite contribution to the integral
in the numerator.
This establishes the second property of $A_1$.
\subsection{$A_1$ term---enforcement of the Born rule}
\label{Subsection:Born_rule}
In this subsection we must be clear about those quantities that depend on $t_{\rmi}$, so we will revert to the notation $\psi_j(t, \vec x; t_{\rmi})$ introduced in \eqref{define_tau} and \eqref{define_psi_basis}. Then the normal-mode expansion of $\psi$ in \eqref{expansion_in_normal_modes} must be understood as
\begin{equation*}
\psi(t, \vec x ; t_{\rmi}) = \sum_j C_j(t ; t_{\rmi}) \, \psi_j(t, \vec x; t_{\rmi}) ,
\end{equation*}
reflecting the fact that the evolution of the system, as expressed by the coefficients $C_j$ in the expansion, also depends on $t_{\rmi}$. [We will henceforth write the derivative $C_j^{\, \prime}$ in identity \eqref{identity_pi-m} as a partial derivative.]
Then the proportion (``branching ratio'') of decays of a wavefunction given by \eqref{expansion_in_normal_modes} to the single state $\psi_j$ must be the initial weight 
\begin{equation}
\label{initial_weights}
Y_j \equiv |C_j(t_{\rmi};t_{\rmi})|^2
\end{equation}
of that state.

We wish to study the statistics of the experimental outcomes as the starting time is varied. 
If all modes present had the same frequency, we could average the starting time $t_{\rmi}$ over one period of oscillation. Since nontrivial systems will have multiple frequencies, that average is complicated, and we choose instead to average $t_{\rmi}$ over an interval long compared to the periods of all the modes present.
To denote the average on $t_{\rmi}$, we will use an overbar and drop the argument $t_{\rmi}$, thus for an arbitrary function $F$,
\begin{equation*}
\overline{F(t, a)} \equiv \lim_{T \to \infty} \frac{1}{T} \int^t_{t-T} dt_{\rmi} \, F(t, a; t_{\rmi})
\end{equation*}
where the optional argument $a$ stands for any set of independent variables besides $t$ and $t_{\rmi}$.

Now we expect that the term $A_2$ in the VP will cause the system to decay to a single state, that is,
\begin{equation}
\label{appearance_of_final_decay}
\lim_{t \to \infty} \left| C_j(t; t_{\rmi}) \right|^2 = \left\{ {
\begin{array}{rl}
1  & (j=k) \\
0  & (j \neq k)
\end{array}
} \right.
\end{equation}
for some $k$ (which, we presume, depends on the choice of $t_{\rmi}$).
Then, in a large number of ``identically prepared'' experiments, the fraction of outcomes in state $j$ is just the average over values of $t_{\rmi}$ of the left-hand side of \eqref{appearance_of_final_decay}.
Then the Born rule can be expressed precisely\footnote{In the general case,
where the decay is to a set of degenerate states, we replace $|C_j|^2$ everywhere it appears in equation \eqref{Born_rule} by the sum of that quantity over the degenerate set.}
as
\begin{equation}
\label{Born_rule}
Y_j = \overline{\lim_{t \to \infty} | C_j(t) |^2} .
\end{equation}
This is the ``branching ratio'' property that our theory must reproduce.

Because $\{\psi_j(t)\}$ are a complete set,
we can write
\begin{equation*}
\begin{split}
\left\langle \psi \right| \mathcal{D}^\dagger \mathcal{D} \left| \psi \right\rangle_t
&= \sum_j \left\langle \psi \right| \mathcal{D}^\dagger \left| \psi_j \right\rangle_t
\left\langle \psi_j \right| \mathcal{D} \left| \psi \right\rangle_t \\
&= \sum_j \left| T_j \right|^2
\end{split}
\end{equation*}
where
\begin{equation*}
T_j(t; t_{\rmi})
\equiv
\left\langle \psi_j \right| \mathcal{D} \left| \psi \right\rangle_t .
\end{equation*}
Thus the minimization of $A_1$ tends to make
$\left| T_j \right|$
as small as possible for every energy eigenstate $\psi_j$, at every time $t$.

If $A_1$ were the only term in the variational principle, it would attain a value of zero by making every $T_j$ vanish. But it is in competition with $A_2$, which is trying to cause a superposition of modes to decay to a single mode, which requires some coefficients $C_j$ to have nonzero time derivatives, which [as we see from \eqref{A_1_as_sum_of_C_prime_sqd}] prevents $A_1$ from going all the way to zero. So we will have to analyze that competition to determine how small the $T_j$'s will be. That is beyond the scope of this paper. Instead, let us study the effect of allowing one $T_j$ to vanish completely; this should approximate the effect of its being small but nonzero.

We therefore set
\begin{equation}
\label{begin_branching_ratio}
T_j = \left\langle \psi_j \right| \mathcal{D} \left| \psi \right\rangle_t = 0
\end{equation}
for any $j$. Then we can use identity \eqref{identity_pi-m} to write
\begin{equation}
\label{A_2_analysis_2}
\begin{split}
0
&= C_j^* \sum_k \frac{\partial C_k}{\partial t} \left\langle \psi_j \right| \gamma^{\,0} \left| \psi_k \right\rangle_t \\
&= \frac{\partial}{\partial t} \left(|C_j|^2 \right) \left\langle \psi_j \right| \gamma^{\,0} \left| \psi_j \right\rangle_t
+ \sum_{k \ne j}
C_j^* \frac{\partial C_k}{\partial t} \left\langle \psi_j \right| \gamma^{\,0} \left| \psi_k \right\rangle_t
\end{split}
\end{equation}

Further progress depends upon understanding the matrix elements $\left\langle \psi_j \right| \gamma^{\,0} \left| \psi_k \right\rangle_t$. If the form \eqref{choice_of_gammas} is used in the nonrelativistic case, it is easily seen from the wavefunctions \eqref{order_magnitude_positive} and \eqref{order_magnitude_negative} and the orthonormality relation \eqref{orthonormality_psi} that if $E_j$ and $E_k$ have the same sign,
\begin{equation*}
\left\langle \psi_j \right| \gamma^{\,0} \left| \psi_k \right\rangle_t = \delta_{jk} \, \sgn{E_j} + O \left( \tfrac{p^2}{m^2} \right).
\end{equation*}
If they have different signs, so that $\lvert E_j - E_k \rvert \approx 2m$, then $\left\langle \psi_j \right| \gamma^{\,0} \left| \psi_k \right\rangle_t$ is $O \left( \frac{p}{m} \right)$ and, by Eq. \eqref{define_psi_basis}, has the value
\begin{equation*}
\left\langle \psi_j \right| \gamma^{\,0} \left| \psi_k \right\rangle_t
= \left\langle \chi_j \right| \gamma^{\,0} \left| \chi_k \right\rangle_{t-t_{\rmi}}
\exp \left[ i \int_0^t \Delta E_{jk}(t^\prime - t_{\rmi}) \, dt^\prime \right],
\end{equation*}
in which we have introduced the shorthand
\begin{equation*}
\Delta E_{jk}(\tau) \equiv E_j(\tau) - E_k(\tau) \, .
\end{equation*}
This is a rapidly oscillating function of $t$ (with an angular frequency of the order of $2m$), to wit, zitterbewegung.

So, keeping terms to $\Or(p/m)$ in \eqref{A_2_analysis_2}, we see that
\begin{multline}
\label{A_2_analysis_3}
\frac{\partial}{\partial t} \left( \left|C_j(t; t_{\rmi}) \right|^2 \right) = \\
-\sgn{E_j} \mspace{-6mu} \sum_{\substack{k \ne j \\ E_j E_k < 0}}
\mspace{-6mu} C_j^*(t; t_{\rmi}) \frac{\partial C_k(t; t_{\rmi})}{\partial t}
\left\langle \chi_j \right| \gamma^{\,0} \left| \chi_k \right\rangle_{t-t_{\rmi}}
\exp \left[ i \int_0^t \Delta E_{jk}(t^\prime - t_{\rmi}) \, dt^\prime \right] .
\end{multline}
Thus $|C_j|^2$ is governed by a sum of zitterbewegung terms, causing it to vary on a very short timescale and in a way that depends on $t_{\rmi}$. We will ensure by the action of $A_2$ that the wavefunction will decay to a single state [Eq. \eqref{appearance_of_final_decay}], but it appears that the zitterbewegung terms in \eqref{A_2_analysis_3} will determine in which state it ends up, and the different possible outcomes will arise from different choices of $t_{\rmi}$. (We will soon demonstrate that this is true.) This is the ``hidden variable'' behavior we predicted.

Note also that the terms determining the evolution of $|C_j|^2$ are of the order of the momentum $p$ of the individual eigenstates. This linear dependence on momentum is consistent with our earlier argument about how the zitterbewegung terms must scale in order to determine the outcome of the wavefunction decay.

Now we integrate the last equation and use the initial weights \eqref{initial_weights}:
\begin{equation}
\label{integrate_weights}
\left| C_j(t; t_{\rmi}) \right|^2
= Y_j + \int_{t_{\rmi}}^t \frac{\partial}{\partial t^\prime} \left( \left|C_j(t^\prime; t_{\rmi}) \right|^2 \right) \,
dt^\prime
\end{equation}
With a little effort we find that
\begin{multline}
\label{C_j_sqd_integrated}
\left| C_j(t; t_{\rmi}) \right|^2
\approx
\,\, Y_j + i \, \sgn{E_j} \\
\times \mspace{-6mu} \sum_{\substack{k \ne j \\ E_j E_k < 0}}
\left\{
\frac {C_j^*(t; t_{\rmi}) \,\, \frac{\partial C_k(t; t_{\rmi})}{\partial t}
\, \left\langle \chi_j \right| \gamma^{\,0} \left| \chi_k \right\rangle_{t-t_{\rmi}}} 
{\Delta E_{jk} (t-t_{\rmi})}
\,\exp \left[ i \int_0^t \Delta E_{jk}(t^\prime - t_{\rmi}) \, dt^\prime \right]
\right. \\
-
\left.
\frac {C_j^*(t_{\rmi}; t_{\rmi}) \,\, \frac{\partial C_k(t; t_{\rmi})}{\partial t} |_{t=t_{\rmi}}
\, \left\langle \chi_j \right| \gamma^{\,0} \left| \chi_k \right\rangle_0} 
{\Delta E_{jk} (0)}
\, \exp \left[ i \int_0^{t_{\rmi}} \Delta E_{jk}(t^\prime - t_{\rmi}) \, dt^\prime \right]
\right\} .
\end{multline}
(We give mathematical details in the appendix.)

Next we need to take the $t\to \infty$ limit. But we know that due to \eqref{appearance_of_final_decay}, the partial derivative $\partial C_k / \partial t$ goes to zero in that limit, and the other factors in the first term in curly brackets are bounded, so that term drops out:
\begin{multline*}
\lim_{t \to \infty} \left| C_j(t; t_{\rmi}) \right|^2
\approx 
\,\, Y_j
\\
- i \, \sgn{E_j}
\mspace{-9mu} \sum_{\substack{k \ne j \\ E_j E_k < 0}}
\frac {C_j^*(t_{\rmi}; t_{\rmi}) \,\, \frac{\partial C_k(t; t_{\rmi})}{\partial t} |_{t=t_{\rmi}}
\, \left\langle \chi_j \right| \gamma^{\,0} \left| \chi_k \right\rangle_0} 
{\Delta E_{jk} (0)}
\,\, \exp \left[ i \int_0^{t_{\rmi}} \Delta E_{jk}(t^\prime - t_{\rmi}) \, dt^\prime \right]
.
\end{multline*}
Finally, we observe that none of the factors in the numerator and denominator of the fraction (with the possible exception of the partial derivative) actually depend on the start time $t_{\rmi}$; for instance, $C_j^*(t_{\rmi}; t_{\rmi})$ is one of the quantities that would be set by the experimental design. Then when we average on $t_{\rmi}$, the complex exponential averages to zero (to good approximation). The result is the Born rule \eqref{Born_rule}, which is what we set out to prove.

Since the statistical distribution of outcomes (and thus all possible outcomes) resulted from an average over experimental start times $t_{\rmi}$, it follows that the different outcomes were determined by the choice of $t_{\rmi}$. 
This establishes the role of the start time of the experiment relative to the oscillations of the wavefunction (or equivalently, the wavefunction's phase at a given time relative to the experiment) as the necessary hidden variable that determines the outcome.
Our derivation shows that zitterbewegung is the mechanism by which the choice of $t_{\rmi}$ determines the outcome. 

Note that this derivation is approximate. The supposition in \eqref{begin_branching_ratio} that $T_j$ is exactly zero is not accurate; 
as we pointed out, its size will actually depend on how well the $A_2$ term fares in its competition with $A_1$. Also, the analysis after
equation \ref{begin_branching_ratio} includes some approximations. Therefore the appropriate conclusion is that the Born rule is an approximate law. This suggests that natural phenomena may disagree with the Born rule under some circumstances, and by some (presumably small) amounts, that
could be predicted by our theory. This may constitute an opportunity for a feasible experimental test of the VP. 
(See the last paragraph in subsection \ref{Subsection:Constraints}.)

\subsection{$A_2$ term---wavefunction collapse}
In order to construct $A_2$ we begin by constructing an expression for (squared) position uncertainty $\delta x^2$. It would seem natural to define it by
\begin{equation}
\label{simple_deltax2}
\delta x^2
\equiv \, \expectationfit{\expectationfit{\lvert \vec x_1 - \vec x_2 \rvert ^2}}_2
\, ,
\end{equation}
where the expectation of a two-coordinate operator $\expectation{\expectation{\mathcal{O}(x_1,x_2)}}_2$ is defined by analogy to 
$\expectationfit{\expectationfit{\mathcal{O}}}_1$.\footnote{
\label{footnote_re_delta_x2}
Note that the expression in \eqref{simple_deltax2} is twice another common expression for the squared position uncertainty:
${\expectationtwo{\lvert \vec x_1 - \vec x_2 \rvert ^2}}_2 
= {\expectationtwo{\lvert \vec x_1 \rvert ^2}}_1 - 2 {\expectationtwo{\vec x_1 \cdot \vec x_2}}_2 + {\expectationtwo{\lvert \vec x_2 \rvert ^2}}_1 
= {\expectationtwo{\lvert \vec x_1 \rvert ^2}}_1 - 2 {\expectationtwo{\vec x_1}}_1 \! \cdot \! {\expectationtwo{\vec x_2}}_1 + {\expectationtwo{\lvert \vec x_2 \rvert ^2}}_1 
= 2 \left[ {\expectationtwo{\lvert \vec x_1 \rvert ^2}}_1 - ({\expectationtwo{\vec x_1}}_1)^2 \right] 
= 2 \, {\expectationtwo{\lvert \vec x_1 - {\expectationtwo{\vec x_1}}_1 \rvert ^2}}_1$.
However, we are about to redefine ${\expectationtwo{\mathcal{O}(x_1,x_2)}}_2$ in a way that will prevent ${\expectationtwo{\vec x_1 \cdot \vec x_2}}_2$ from being factored in this way except in particular special cases, so our uncertainty expression will in general not be simply related to ${\expectationtwo{\lvert \vec x_1 - {\expectationtwo{\vec x_1}}_1 \rvert ^2}}_1$.
}
We need to modify this definition in two ways. First, we replace the operator within the angle brackets by the relativistically covariant expression 
$-(x_1^\mu - x_2^\mu)(x_{1\mu} - x_{2\mu})$.
(We will use the summation convention for repeated Greek indices, which run from 0 to 3.)

The second modification is motivated by our intention that the (squared) wavefunction uncertainty
$\delta x^2 \delta p^2$ should be meaningful at, or at least near, a given instant in time, so our expression for it must couple positions and momenta that are ``at the same time'' in some sense.
Since the property of simultaneity depends on choice of reference frame, we will instead require that the spacetime locations sampled in computing $\delta x^2 \delta p^2$ be spacelike separated. Therefore we will include within integrands some function $f(x_1-x_2)$ that vanishes whenever $x_1$ and $x_2$ are timelike separated.
The simplest choice of $f$ is of course
\begin{equation}
\label{simple_f}
f(z) \equiv u \left( -z^\mu z_\mu \right)
\end{equation}
where $u$ is the unit step (Heaviside) function, but many other forms are possible.

With these changes, we define the expectation of a two-coordinate one-particle operator as
\begin{equation}
\label{define_<O>_2}
\expectationfit{ \expectationfit{ \mathcal{O}_2 } }_2 \,
\equiv \, 
\frac
{\int d^4 x_1 \, d^4 x_2 \,\, \psi^\dagger(x_1) \, \psi^\dagger(x_2) \, \mathcal{O}_2(x_1,x_2) \, \psi(x_1) \, \psi(x_2) \, f(x_1-x_2)}
{\int d^4 x_1 \, d^4 x_2 \,\, \psi^\dagger(x_1) \, \psi^\dagger(x_2) \,                                                \psi(x_1) \, \psi(x_2) \, f(x_1-x_2)}
\, .
\end{equation}
In this definition, the subscript $2$ attached to $\mathcal{O}$ is a reminder that it depends on two arguments; the subscript $2$ on the triangular brackets is to distinguish the expectation formula defined here, which applies to two-argument operators, from expectations defined in \eqref{define_<O>} and in other definitions below.

We are familiar with the utility in SQM of the three-dimensional inner product integral
\begin{equation}
\label{3D_orthogonality_integral}
\int d^3 x \: \psi_1^\dagger(t, \vec x) \, \psi_2(t, \vec x) \, ,
\end{equation}
which is central to, for instance, orthonormality relations needed for the construction and use of basis sets. We would like the relativistic theory we are developing to have similar properties in appropriate limiting cases. In particular, we note that if the products $\psi^\dagger(x_1) \, \psi^\dagger(x_2) \, \psi(x_1) \, \psi(x_2)$ and $\psi^\dagger(x_1) \, \psi^\dagger(x_2) \, \mathcal{O}(x_1,x_2) \, \psi(x_1)$ $\psi(x_2)$ are constant in time, then the temporal integrations on $x_2^{\,0}$ in \eqref{define_<O>_2} can be performed first. If we choose the unit step function \eqref{simple_f} for $f$, the inner integral is
\begin{equation*}
\int dx_2^{\,0} \, f(x_1-x_2)
= 2 \, \lvert \vec x_1 - \vec x_2 \rvert \, ,
\end{equation*}
expressing the fact that at the location $\vec x_2$, the time interval $\Delta x_2^{\,0}$ over which $(x_2^{\,0}, \vec x_2)$ is spacelike separated from $(x_1^{\,0}, \vec x_1)$ is proportional to the spatial separation $\lvert \vec x_1 - \vec x_2 \rvert$. As we prefer not to give greater weight to greater separation distances in expectation calculations such as \eqref{define_<O>_2}, we might take instead of \eqref{simple_f} the form
\begin{equation}
\label{trial_f}
f(z) \equiv \frac {u \left( -z^\mu z_\mu \right)} {2 \, \lvert \vec z \rvert}
\, .
\end{equation}
Then
\begin{equation}
\label{integrate_final_f}
\int dx_2^{\,0} \, f(x_1-x_2) = 1
\, ,
\end{equation}
so the numerator and denominator in \eqref{define_<O>_2} factor into products of orthogonality integrals like \eqref{3D_orthogonality_integral}---a useful property, as we will see in due time. Unfortunately, this form of $f$ is not relativistically covariant, but if we choose instead
\begin{equation*}
f(z)
\equiv
\frac
{u \left( -z^\mu z_\mu \right)}
{\pi \sqrt{-z^\mu z_\mu}}
\, ,
\end{equation*}
we find that \eqref{integrate_final_f} is still satisfied.

We might in this way define position uncertainty as
\begin{equation}
\label{define_delta_x_2}
\delta x^2
\equiv \, \expectationfit{\expectationfit{-(x_1^\mu - x_2^\mu)(x_{1\mu} - x_{2\mu})}}_2
\end{equation}
and momentum uncertainty as
\begin{equation}
\label{define_delta_p_2}
\delta p^2
\equiv \,
\expectationfit{\expectationfit{
- \left[ p_1^\mu(x_1) - p_2^\mu(x_2) \right] \left[ p_{1\mu}(x_1) - p_{2\mu}(x_2) \right]
}}_2
\, ,
\end{equation}
and define $A_2$ as the product of $\delta x^2$ and $\delta p^2$, but that would give cross terms composed of non-conjugate variable pairs, such as 
$\delta y^2 \,\delta p_z^2$. Instead, let us use the combination $(\delta x^\mu \, \delta p_\mu)^2$ within the angle brackets. It will also give unwanted cross terms 
(e.g., $\delta x^{\mu=1} \, \delta p_{\mu=1} \delta x^{\nu=2} \, \delta p_{\nu=2}$), but they will not be quadratic and should therefore not contribute to expectation values. This leads us to define
\begin{equation}
\delta x^2 \delta p^2
\equiv
\left\{ (x_1^\mu - x_2^\mu) \, [p_{3 \mu}(x_3) - p_{4 \mu}(x_4)] \right\}^2
\end{equation}
[where the LHS is simply the notation for a new operator, and not the product of
\eqref{define_delta_x_2} and \eqref{define_delta_p_2}]
and
\begin{equation}
\label{define_A_2}
A_2  =
\expectationfit{ \expectationfit{ \delta x^2 \delta p^2 }}_4
\, ,
\end{equation}
in which $\expectation{\expectation{\mathcal{O}_4}}_4$ is an extension of the expectation defined in \eqref{define_<O>_2} to operators depending on four spacetime points:
\begin{multline}
\expectationfit{\expectationfit{\mathcal{O}_4}}_4 \,
\equiv
\\
\shoveleft{
\bigg[
\int d^4 x_1 \, d^4 x_2 \, d^4 x_3 \, d^4 x_4 \;
\psi^\dagger(x_1) \,\psi^\dagger(x_2) \,  \psi^\dagger(x_3)\,  \psi^\dagger(x_4) \,
\mathcal{O}_4(x_1, \ldots x_4)
}
\\
\shoveright{
\psi(x_1) \, \psi(x_2) \, \psi(x_3) \, \psi(x_4) \, f( \{x_k-x_l : 1\! \le \! k \! < \! l \! \le \! 4\} ) 
\bigg]
}
\\
\shoveleft{
\bigg[ \int d^4 x_1 \, d^4 x_2 \, d^4 x_3 \, d^4 x_4 \;
\psi^\dagger(x_1) \, \psi^\dagger(x_2) \, \psi^\dagger(x_3) \, \psi^\dagger(x_4)
}
\\
\psi(x_1) \, \psi(x_2) \, \psi(x_3) \, \psi(x_4) \,
f( \{x_k-x_l : 1\! \le \! k \! < \! l \! \le \! 4\} )
\bigg] ^{-1}
\end{multline}
and the function $f( \{x_k-x_l : 1\! \le \! k \! < \! l \! \le \! 4\} )$ enforces the spacelike separation of all four points:
\begin{equation}
\label{final_f_4pt}
f(x_1-x_2, x_1-x_3, \ldots x_3-x_4)
\equiv
\frac
{\prod^3_{k=1} \prod^4_{l=k+1} u \left[-(x_k^\mu - x_l^\mu)(x_{k\mu} - x_{l\mu}) \right]}
{W(x_1-x_2, x_1-x_3, \ldots x_3-x_4)}
\, .
\end{equation}
Here the weight function $W$ must be chosen so that $f$ satisfies the four-point extension of \eqref{integrate_final_f}:
\begin{equation}
\label{integrate_final_f_4pt}
\int dx_2^{\,0} \int dx_3^{\,0} \int dx_4^{\,0} \,\, f(x_1-x_2, x_1-x_3, x_1-x_4, x_2-x_3, x_2-x_4, x_3-x_4) = 1
\, .
\end{equation}
As we saw in \eqref{trial_f} for the expectation of a two-point operator, a trivial solution is
\begin{multline*}
W(x_1-x_2, x_1-x_3, \ldots x_3-x_4)
=
\\
\int dx_2^{\,0} \int dx_3^{\,0} \int dx_4^{\,0} \,\,
\prod^3_{k=1} \, \prod^4_{l=k+1} u \left[-(x_k^\mu - x_l^\mu)(x_{k\mu} - x_{l\mu}) \right]
\, ,
\end{multline*}
which is unfortunately not covariant because it is a function of 
$\lvert \vec x_1- \vec x_2 \rvert, \lvert \vec x_1 - \vec x_3 \rvert, \ldots \lvert \vec x_3 - \vec x_4 \rvert$ but not 
$x^{\,0}_1-x^{\,0}_2, x^{\,0}_1-x^{\,0}_3, \ldots x^{\,0}_3-x^{\,0}_4$.
We conjecture that a covariant weight function $W$ satisfying \eqref{integrate_final_f_4pt} exists, and will proceed to use it without determining its form.
\subsection{$N$-particle version of the variational principle}
When no measurement is being performed, an isolated system obeys the usual SQM wave equation.
On the other hand, if a measurement is being made,
$\mathcal{R}$ includes the wavefunctions of both the system and (some part of) the measuring apparatus,
for the reason given in the discussion of $A_2$ in subsection \ref{Subsection:VP_terms}.
Then we must generalize the variational principle \eqref{variational_principle} to describe the set of all the particles in $\mathcal{R}$. Let those particles be labeled with the subscript $n$, where for instance $n=1$ might be the ``system'' being measured, and $n>1$ are particles of the apparatus. Suppose for simplicity that all $N$ particles are distinguishable and have spin $\frac{1}{2}$. Then we generalize equations \eqref{define_A_1} and \eqref{define_<O>} to
\begin{equation}
A_1  =
\sum_{n=1}^N
\expectationfit{\expectationfit{\mathcal{D}_n^\dagger \mathcal{D}_n}}_1
\end{equation}
and
\begin{multline}
\label{define_<O>_1_general}
\expectation{\expectation{\mathcal{O}}}_1 \,
\equiv
\\
\frac
{\int \left( \prod_n d^4 x_n \right) \,\, \psi^\dagger(x_1,x_2,x_3,\ldots,x_N) \, \mathcal{O}(x_1,x_2,x_3,\ldots,x_N) \, \psi(x_1,x_2,x_3,\ldots,x_N)}
{\int \left( \prod_n d^4 x_n \right) \,\, \psi^\dagger(x_1,x_2,x_3,\ldots,x_N) \, \psi(x_1,x_2,x_3,\ldots,x_N)}
\, .
\end{multline}
For operators depending on four points per particle,
we generalize equation \eqref{define_A_2} to
\begin{equation}
A_2  =
\Bigg\langle \Bigg\langle
\sum_{n=1}^N
\left\{ (x_{n1}^\mu - x_{n2}^\mu) \, [p_{n3 \mu}(x_3) - p_{n4 \mu}(x_4)] \right\}^2
\Bigg\rangle \Bigg\rangle_4 \,
\, ,
\end{equation}
in which the four-point expectation $\expectation{\expectation{\,\,}}_4$ is defined as
\begin{multline}
\expectationfit{\expectationfit{\mathcal{O}_4}}_4 \,
\equiv \\
\shoveleft{
\Bigg[
\int \left( \prod_{n=1}^N \, \prod_{k=1}^4 d^4 x_{nk} \right) \,\,
\left( \prod_{k=1}^4 \psi^\dagger(x_{1k}, \ldots x_{Nk}) \right)
}
\mathcal{O}(x_{11}, x_{12}, x_{13}, x_{14}, x_{21}, \ldots, x_{N4})
\\
\shoveright{
\left( \prod_{k=1}^4 \psi(x_{1k}, \ldots x_{Nk}) \right)
\prod_{n=1}^N \, f( \{x_{nk}-x_{nl} : 1\! \le \! k \! < \! l \! \le \! 4\} )
\Bigg]
}
\\
\Bigg[
\int \left( \prod_{n=1}^N \, \prod_{k=1}^4 d^4 x_{nk} \right) \,
\left( \prod_{k=1}^4 \lvert \psi(x_{1k}, \ldots x_{Nk}) \rvert ^2 \right)
\prod_{n=1}^N \, f( \{x_{nk}-x_{nl} : 1\! \le \! k \! < \! l \! \le \! 4\} )
\Bigg]
^{-1}
\end{multline}
in which the notation $x_{n1}, \ldots x_{n4}$ signifies four different spacetime coordinates for particle $n$, and $f$ has the form \eqref{final_f_4pt}.

The $N$-particle version of the VP has recently been applied by this author \cite{Harrison_FQMT}
to the electron two-slit experiment \cite{Jonsson,Feynman}, including Wheeler's delayed-choice variant. \cite{Wheeler_1979}

\subsection{Comparison of the VP to the design constraints}
The VP we have constructed satisfies most or all of the constraints listed in subsection \ref{Subsection:Constraints}.
Nothing in the theory prevents us from understanding it as a description of matter waves themselves, so the first property is satisfied.
Since the VP is the sum of terms providing for both state reduction, when that is called for, and behavior consistent with
the SQM wave equation otherwise, it has properties 2-4.
We have shown that it approximately satisfies the Born rule, and conjecture that the approximation is good enough that the
experimental record does not contradict it; this is just the fifth constraint. That analysis also confirmed that the phase of the wavefunction,
as exhibited in the phenomenon of zitterbewegung, plays the role of the hidden variable, as predicted by constraint 6.

As we will discuss in subsection \ref{Determinism}, we are unable to say whether the new theory is deterministic, which was property 7.
However, the form of the integrals makes it clear that properties 8-12 are satisfied.
We expect property 13 to hold as well; certainly we have not constructed different forms of the theory for microscopic and macroscopic domains.

\section{Feasibility of experimental tests of the theory}
It seems plausible that this theory could be tested experimentally. One promising avenue is the decay process, as opposed to the collapse favored by SQM. We have noted that SQM embargoes any information derived from a measurement in less than the time 
$\Delta t \approx 1/\Delta E$; but we have not found such a limitation necessary in our theory. It would be interesting to make measurements within that time interval to see if the decay process could be detected.

Another possibility is to look for correlations between events close together in time. 
If a system really evolves deterministically, depending on the hidden variable $t_{\rmi}$, 
then the correlation between two measurements made in rapid succession
on the same system may show evidence of that.
However, the interesting content of the correlation function may decay or oscillate on a timescale comparable to the time 
$T_{jk}$ given by \eqref{very_short_period},
so the required timing precision may be unattainable with current technology.

We have also admitted that our derivation of the Born rule includes some approximations. If in fact the theory satisfies that rule only approximately, then the deviations from the exact rule constitute predictions that could be tested experimentally,
as we suggested at the end of subsection \ref{Subsection:Born_rule}.

\section{New perspectives}
Quantum mechanics has challenged physicists' intuition since its inception, because it is understood to operate in ways unlike any other physical theory. Although we continue to embrace many such ideas, such as the intrinsic nonlocality of nature, we have attempted to overturn some that we considered particularly objectionable, such as the special but ill-defined treatment of measurements. We have certainly not succeeded in rewriting quantum mechanics in an orderly, conventional form like, for instance, classical electrodynamics or even special relativity (nor did we expect to). In fact, at this stage of our understanding, we appear to have introduced some new enigmas. Nevertheless, we have found a natural way to assimilate both nonlocality and the measurement process into a variational principle that reduces to the Dirac equation under appropriate conditions. Although there are still important mysteries about how to put the pieces together, the elements of our variational principle appear to encapsulate the essentials of a theory that may have advantages compared to those that have been explored and accepted up to now.

We shall at this point take note of the perspective this theory gives on some well-known issues.

\subsection{The ``uncertainty principle''}
We will observe here that (at least) two different ideas are commonly referred to as ``the (Heisenberg) uncertainty principle.'' One is the idea of complementarity, that there are pairs of ``complementary variables'' for which the product of the two uncertainties has a minimum value of the order of $\hbar$. Since we are taking $\hbar \equiv 1$, those inequalities
\begin{align*}
\Delta x \, \Delta p &\ge 1 \, , &\Delta t \, \Delta E &\ge 1
\end{align*}
are equivalent to mathematical relations familiar (in an order-of-magnitude sense) from Fourier analysis:
\begin{align*}
\Delta x \, \Delta k &\ge 1 \, , &\Delta t \, \Delta \omega &\ge 1 \, .
\end{align*}
Clearly complementarity is a valid principle, expressing fundamental mathematical properties of waves.

On the other hand, the term ``uncertainty principle'' is also often used to refer to the idea that the measurement process is \textit{intrinsically} random, so the outcome of a measurement is uncertain. Our theory rejects that uncertainty principle, proposing instead that the result of a measurement depends on the phase of the wavefunction.

\subsection{Time-reversal invariance}
We observed earlier that the collapse process in SQM is time-reversal-dependent, as it converts multiple states in the past into a single state in the future. We note that our variational principle, like other fundamental laws, has no preference for either direction of time.

The time reversal dependence of SQM is actually more subtle than is often appreciated. The state of a system before a measurement may be a superposition only in terms of the operator, or set of physical fields, that will be imposed to make the measurement.
For instance, an electron with spin $\frac{1}{2}$ in the $z$ direction exists in a single eigenstate of the operator $S_z$. If we then consider its spin in the $x$ direction, we describe it as being in a superposition of two spin states, $\pm \frac{1}{2}$. But it has not changed its state; we have simply chosen a different basis set in which to describe it. After the measurement, when it is in a single eigenstate of the $S_x$ operator, we could equally well describe it as in a superposition of two eigenstates of $S_z$. So this case is really time-reversal invariant, even according to SQM.

This idea stands out more clearly when we compare the perspectives of two observers ``traveling'' in opposite directions through time. Suppose that at $t=0$ we measure $S_x$ by quickly switching the alignment of an imposed magnetic field from the $z$ to the $x$ direction. An observer traveling ``forward'' in time (that is, in the direction we sense as forward) would say that the single state with $S_z = \frac{1}{2}$, understood as two states with $S_x = \pm \frac{1}{2}$, collapsed at $t=0$ to a single state, say $S_x = \frac{1}{2}$. An observer coming from our future toward our past would observe the single state with $S_x = \frac{1}{2}$ and note that at $t=0$ the magnetic field changes from the $x$ to the $z$ direction. He would conclude that we had measured the $z$ component of its spin, causing the two states $S_z = \pm \frac{1}{2}$ to collapse to the single one with $S_z = \frac{1}{2}$. Therefore both observers would succeed in interpreting events as consistent with SQM. For this thought experiment, the predictions of SQM are actually symmetric in time.

The predictions of our theory are similar in this case, except that the transition of the wavefunction from an eigenstate of $S_z$ to an eigenstate of $S_x$ occurs smoothly around $t=0$. Since the variational principle has no sensitivity to the direction of time, that transition is presumably symmetric about $t=0$. This is an example of the nonlocality in time referred to earlier. Because the transition begins before the instant at which the field alignment changed, this is a violation of causality in the usual sense, although others \cite{Bennett_1,Bennett_2} have pointed out that quantum mechanics can violate causality. (The thought experiment does have a flaw, however, in that the switching of the field alignment cannot really be instantaneous.)

Now let us modify that thought experiment somewhat. Suppose that at $t=0$ we turn off the field in the $z$ direction, and then cause the $x$-aligned field to appear at $t=1$. Our two observers (both trained in SQM) will draw conclusions consistent with SQM as before. However, their conclusions will be inconsistent with each other, because the first observer will maintain that the system stayed in the state with $S_z = \frac{1}{2}$ until we measured it at $t=1$. The second observer will regard the appearance (from his perspective) of the $z$-aligned field at $t=0$ as our measurement, so he will conclude that for $0<t<1$ the electron was still in the $S_x = \frac{1}{2}$ state that he had observed at times $t>1$. This is a case in which the predictions of SQM vary with the direction of time.

On the other hand, our theory will maintain time-reversal invariance. The evolution of the wavefunction during the period of interest will be determined by minimizing the integral (over time and space) of the appropriate functional, as has been described. Whatever the state of the electron between $0$ and $1$, it will be the same for both observers.

\subsection{Is the new theory deterministic?}
\label{Determinism}
We hoped to produce a deterministic theory; in principle, that should be possible, because we have a hidden variable. However, it is not yet clear to us how nature might solve the variational principle. 
The wavefunction over an entire region of spacetime is available to be varied. (Actually, all of spacetime could theoretically be involved. But any properly conducted experiment must be isolated from unwanted influences, so there must be spatial bounds on the region that must be considered. In addition, there must be a start time at which the state of the system is well-defined and a stop time at which the results are unambiguous, so the relevant region of spacetime is completely bounded.) Does nature search the entire available phase space and find the solution which gives an absolute minimum in the variational principle? Or is a local minimum sufficient? If a local minimum is sought, how is the search conducted? If multiple local minima are available (or nearby, if nearness in phase space is relevant), how is a single one chosen?

Until we understand more of these issues, we cannot say whether the theory is deterministic.

\section{Summary and conclusions}
SQM explains nature in terms of a wave equation which (whether the Schr\"odinger, Dirac, or Klein-Gordon equation) is linear and local, in spite of the fact that nature is clearly nonlinear and nonlocal. The wave equation is supplemented by a ``collapse'' process that is nonlinear, nonlocal and is usually understood to be time-reversal-dependent. We are suspicious of the adequacy of the wave equation and the plausibility of the collapse process, and the criteria that determine which of the two processes governs at any given instant seem to us to range from vague to unbelievable. Accordingly, we have set out to construct a unified theory that explains both types of phenomena in a natural way.

In order to force the new theory to resemble the SQM wave equation where possible but allow it to make transitions (decays) where it must, we have formulated it as a variational principle. We have found combinations of integrals over space and time that seem to have the desired properties, and also provide the required nonlocality and nonlinearity in a plausible way. One of the terms in the variational principle forces superpositions of states to decay to a single state. The other term prevents unwanted discontinuities in time, provides for the theory to satisfy the Born rule for the distribution of measurement outcomes, and vanishes for Dirac solutions under the right set of circumstances. The theory includes as a hidden variable the phase of the wavefunction, which is manifested physically via beats between modes of different energies, particularly the zitterbewegung between positive- and negative-energy modes.

We are hopeful that the new theory may be testable with currently available technology.

We note that in addition to eliminating the weaknesses of SQM with regard to wavefunction collapse, measurement theory, and nonlocality, the new theory repudiates the intrinsic randomness of nature as understood in SQM. Also, it is completely time-reversal invariant. We suspect that it may be deterministic, but must better understand the application of the variational principle in nature before we can be certain of that.

\appendix

\section*{\label{apdx:integration}APPENDIX: Evaluating the integral in equation \eqref{integrate_weights}}
\setcounter{section}{1}

We will perform the integration in Eq. \eqref{integrate_weights}, for which the integrand is given by Eq. \eqref{A_2_analysis_3}. Since we are considering a nonrelativistic case (that is, one in which all the modal energies are close to $\pm m$), the energy difference 
$\Delta E_{jk}(\tau)$ does not vary much from $\pm 2m$. Then the value of the integral of that quantity in the exponential in \eqref{A_2_analysis_3} is very close to $\pm 2m t$. This suggests that we could profitably make a change of variable from $t$ to
\begin{equation*}
w(t; t_{\rmi}) = \int_0^t \Delta E_{jk}(t^\prime - t_{\rmi}) \, dt^\prime .
\end{equation*}
We also observe that the exponential in \eqref{A_2_analysis_3} varies much more rapidly than the other factors in the summand. Let us therefore define the function $F(w; t_{\rmi})$ by
\begin{equation*}
\Delta E_{jk} (t-t_{\rmi}) \, F[w(t; t_{\rmi}); t_{\rmi}] = C_j^*(t; t_{\rmi}) \, \frac{\partial C_k(t; t_{\rmi})}{\partial t}
\left\langle \xi_j \right| \gamma^{\,0} \left| \xi_k \right\rangle_{t-t_{\rmi}}
\end{equation*}
and note that $F$ varies slowly as a function of its first argument (compared to the exponential).
Then \eqref{A_2_analysis_3} becomes
\begin{equation*}
\frac{\partial}{\partial t} \left(|C_j(t; t_{\rmi})|^2 \right) = 
-\sgn{E_j} \mspace{-6mu} \sum_{\substack{k \ne j \\ E_j E_k < 0}}
\mspace{-6mu} \Delta E_{jk} (t-t_{\rmi}) \, F[w(t; t_{\rmi}); t_{\rmi}] \, e^{iw(t; t_{\rmi})} \, .
\end{equation*}

Now we substitute this into \eqref{integrate_weights}:
\begin{equation*}
\left| C_j(t; t_{\rmi}) \right|^2
= Y_j -\sgn{E_j} \mspace{-6mu} \sum_{\substack{k \ne j \\ E_j E_k < 0}}
\int_{t_{\rmi}}^t \Delta E_{jk} (t^\prime-t_{\rmi}) \, F[w(t^\prime; t_{\rmi}); t_{\rmi}] \, e^{iw(t^\prime; t_{\rmi})} \, dt^\prime \, .
\end{equation*}
But
\begin{equation*}
\frac{\partial w}{\partial t} = \Delta E_{jk}(t - t_{\rmi})
\end{equation*}
so we can change the variable of integration from $t^\prime$ to $w$, thus:
\begin{equation*}
\left| C_j(t; t_{\rmi}) \right|^2
= Y_j - \sgn{E_j} \mspace{-6mu} \sum_{\substack{k \ne j \\ E_j E_k < 0}}
\int_{w(t_{\rmi}; t_{\rmi})}^{w(t; t_{\rmi})} F(w; t_{\rmi}) \, e^{iw} \, dw \, .
\end{equation*}

Now we integrate by parts.
\begin{multline*}
\left| C_j(t; t_{\rmi}) \right|^2
=
Y_j - \sgn{E_j} \mspace{-6mu} \sum_{\substack{k \ne j \\ E_j E_k < 0}}
\left[ -iF(w; t_{\rmi}) \, e^{iw} \Big|_{w=w(t_{\rmi};t_{\rmi})}^{w(t;t_{\rmi})} - \int_{w(t_{\rmi}; t_{\rmi})}^{w(t; t_{\rmi})} \frac{\partial F(w; t_{\rmi})}{\partial w} \, e^{iw} \, dw \right] .
\end{multline*}
But because $F$ is slowly varying, the second term within the brackets (the integral) is small compared to the first, therefore,
\begin{equation*}
\left| C_j(t; t_{\rmi}) \right|^2
\approx Y_j + i \, \sgn{E_j} \mspace{-6mu} \sum_{\substack{k \ne j \\ E_j E_k < 0}}
\left\{ F[w(t;t_{\rmi}); t_{\rmi}] \, e^{iw(t;t_{\rmi})} - F[w(t_{\rmi};t_{\rmi}); t_{\rmi}] \, e^{iw(t_{\rmi};t_{\rmi})} \right\} .
\end{equation*}
Then we substitute the definitions of $F$ and $w$ into this equation to get \eqref{C_j_sqd_integrated}.

\begin{acknowledgements}
This work has been supported by the NNSA ASC program.
The author appreciates support from Jerry Brock, Mark Chadwick and Robert Webster; 
helpful discussions with Salman Habib, Robin Blume-Kohout, Terry Goldman, Howard Brandt, Baolian Cheng and David Sigeti; 
review of an earlier draft by Jean-Fran\c{c}ois Van Huele; 
and detailed discussions with Dale W. Harrison and B. Kent Harrison over a long period of time. 
He is, however, solely responsible for any errors or deficiencies in the work.
\end{acknowledgements}


\begin{thebibliography}{99}
\bibitem{EPR}
Einstein, A., Podolsky, B., Rosen, N.: Can quantum-mechanical description of physical reality be considered complete? Phys. Rev. 47(10),
777-780 (1935)
%
\bibitem{62_yrs}
Miller, A.I. (ed.): NATO Advanced Study Institute on Sixty-Two Years of Uncertainty: Historical, Philosophical, and Physical Inquiries into the Foundations of Quantum Mechanics, Erice, Sicily, Italy, August 5-15, 1989. Series B: Physics, vol. 226. Plenum, New York (1990)
%
\bibitem{Open_Systems}
Breuer, H.-P., Petruccione, F. (eds.): Open Systems and Measurement in Relativistic Quantum Theory, Naples, April 3-4, 1998. Springer, Berlin (1999)
%
\bibitem{Crossroads}
Evans, J., Thorndike, A.S. (eds.): Quantum Mechanics at the Crossroads: New Perspectives from History, Philosophy and Physics. Springer, Berlin Heidelberg New York (2007)
%
\bibitem{Bell_1990}
Bell, J.S.: Against `measurement.' In: Miller, A.I. (ed.) op. cit., pp. 17-31.
%
\bibitem{Bell_Cosmologists}
Bell, J.S.: Quantum mechanics for cosmologists. In: Bell, J.S.: Speakable and Unspeakable
in Quantum Mechanics, Cambridge, 2004, pp. 117-138.
%
\bibitem{Bell_1964}
Bell, J.S.: On the Einstein Podolsky Rosen paradox. Physics (N. Y.) 1, 195-200 (1964)
%
\bibitem{Clauser_Horne_Shimony_Holt}
Clauser, J.F., Horne, M.A., Shimony, A., Holt, R.A.: Proposed experiment to test local hidden-variable theories. Phys. Rev.
Lett. 23, 880-884 (1969)
%
\bibitem{Cramer_1986}
Cramer, J.G.: The transactional interpretation of quantum mechanics. Rev. Mod. Phys. 58, 647-687 (1986)
%
\bibitem{Griffiths_1984}
Griffiths, R. B.: Consistent histories and the interpretation of quantum mechanics. J. Stat. Phys. 36, 219-272 (1984)
%
\bibitem{Griffiths_2002}
Griffiths, R. B.: Consistent resolution of some relativistic quantum paradoxes. Phys. Rev. A 66, 062101 (2002)
%
\bibitem{Thorndike}
Thorndike, A.: What are consistent histories? In: Evans, J., Thorndike, A.S. (eds.) op. cit., pp. 149-157.
%
\bibitem{Pearle_1979}
Pearle, P.: Toward explaining why events occur. Intl. J. Theor. Phys. 18, 489-518 (1979)
%
\bibitem{GRW_1986}
Ghirardi, G.C., Rimini, A., Weber, T.: Unified dynamics for microscopic and macroscopic systems. Phys. Rev. D 34, 470-491 (1986)
%
\bibitem{GPR_1990}
Ghirardi, G.C., Pearle, P., Rimini, A.: Markov processes in Hilbert space and continuous spontaneous localization of systems of
identical particles. Phys. Rev. A 42, 78-89 (1990)
%
\bibitem{Zeh_1970}
Zeh, H.D.: Interpretation of measurement in quantum theory. Foundations of Physics 1, 69-76 (1970)
%
\bibitem{Zurek_1991}
Zurek, W.H.: Decoherence and the transition from quantum to classical. Physics Today 44, 36-44 (1991)
%
\bibitem{Tegmark_Wheeler}
Tegmark, M., Wheeler, J.A.: 100 years of quantum mysteries. Scientific American 284, 68-75 (2001)
%
\bibitem{Decoherence_and_Classical_World}
Joos, E., Zeh, H.D., Kiefer, C., Giulini, D., Kupsch, J., Stamatescu, I.-O.: 
Decoherence and the Appearance of a Classical World in Quantum Theory, 2nd edn. Springer, Berlin Heidelberg New York (2003)
%
\bibitem{Schlosshauer}
Schlosshauer, M., Fine, A.: Decoherence and the foundations of quantum mechanics. In: Evans, J., Thorndike, A.S. (eds.) op. cit., pp. 125-148
%
\bibitem{Bohm_1952a}
Bohm, D.: A suggested interpretation of the quantum theory in terms of ÓhiddenÓ variables. i. Phys. Rev. 85, 166-179 (1952)
%
\bibitem{Bohm_1952b}
Bohm, D.: A suggested interpretation of the quantum theory in terms of ÓhiddenÓ variables. ii. Phys. Rev. 85, 180-193 (1952)
%
\bibitem{Mermin_1993}
Mermin, N.D.: Hidden variables and the two theorems of John Bell. Rev. Mod. Phys. 65, 803-815 (1993)
%
\bibitem{Pearle_1976}
Pearle, P.: Reduction of the state vector by a nonlinear Schr\"odinger equation. Phys. Rev. D 13, 857-868 (1976)
%
\bibitem{Thaller}
Thaller, B.: The Dirac Equation. Texts and Monographs in Physics. Springer, Berlin (1992)
%
\bibitem{BD}
Bjorken, J.D., Drell, S.D.: Relativistic Quantum Mechanics. McGraw-Hill, New York (1964)
%
\bibitem{Strange}
Strange, P.: Relativistic Quantum Mechanics with Applications in Condensed Matter and Atomic Physics. Cambridge (1998)
%
\bibitem{Greiner}
Greiner, W.: Relativistic Quantum Mechanics: Wave Equations, 3rd edn. Springer, Berlin. (2000)
%
\bibitem{Schwinger}
Schwinger, J.: Quantum Mechanics. Springer, Berlin (2001)
%
\bibitem{Harrison_FQMT}
Harrison, A.K.: Calculation of the electron two slit experiment using a quantum mechanical variational principle.
In: Frontiers of Quantum and Mesoscopic Thermodynamics, Jul. 25-30, 2011, Prague, Czech Republic (to appear).
%
\bibitem{Jonsson}
J\"{o}nsson, C.: Elektroneninterferenzen an mehreren k\"{u}nstlich hergestellten feinspalten. Zietschrift f\"{u}r Physik 161, 454-474 (1961)
%
\bibitem{Feynman}
Feynman, R.P.: The Feynman Lectures on Physics, vol. 3. Addison-Wesley, Reading, Massachusetts (1965)
%
\bibitem{Wheeler_1979}
Wheeler, J.A.: Frontiers of time. In: di Francia, N.T. (ed.) Proceedings of the International School of Physics ÔEnrico FermiÕ. Course LXXII. 
Problems in the Foundations of Physics, pp. 395 -492. North Holland, Amsterdam (1979)
%
\bibitem{Bennett_1}
Bennett, C.L.: Evidence for microscopic causality violation. Phys. Rev. A 35, 2409-2419 (1987)
%
\bibitem{Bennett_2}
Bennett, C.L.: Further evidence for causality violation. Phys. Rev. A 35, 2420-2428 (1987)
%
\end{thebibliography}
\end{document}